\documentclass[twocolumn,prb,aps,superbib,tightenlines,floatfix]{revtex4}
\usepackage{amsfonts}
\usepackage{amsmath}
\usepackage{amssymb}
\usepackage{graphicx}

\begin{document}

\title{Resistive switching in nanogap systems on SiO2 substrate}

\author{Jun Yao,$^{1,2}$ Lin Zhong,$^{+,3,4}$ Zengxing Zhang,$^{5}$ Tao He,$^{5}$ Zhong Jin,$^{5}$ Patrick J. Wheeler,$^{6}$ Douglas Natelson,$^{+,3,6}$ and James M. Tour$^{+,4,5}$}
\affiliation{$^{1}$Applied Physics Program, Rice Quantum
Institute; $^{2}$Department of Bioengineering; $^{3}$Department of
Electrical and Computer Engineering; $^{4}$Department of Computer
Science; $^{5}$Department of Chemistry; $^{6}$Department of
Physics and Astronomy, Rice University, 6100 Main St., Houston,
Texas 77005.         $^{+}$ \emph{\textbf{Email addresses
lzhong@rice.edu; natelson@rice.edu; tour@rice.edu}}}

\begin{abstract}
Voltage-controlled resistive switching is demonstrated in various
gap systems on SiO$_{2}$ substrate. The nanosized gaps are made by
different means using different materials including metal,
semiconductor, and metallic nonmetal. The switching site is
further reduced by using multi-walled carbon nanotubes and
single-walled carbon nanotubes. The switching in all the gap
systems shares the same characteristics. This independence of
switching on the material compositions of the electrodes,
accompanied by observable damage to the SiO$_{2}$ substrate at the
gap region, bespeaks the intrinsic switching from post-breakdown
SiO$_{2}$. It calls for caution when studying resistive switching
in nanosystems on oxide substrates, since oxide breakdown
extrinsic to the nanosystem can mimic resistive switching.
Meanwhile, the high ON/OFF ratio ($\sim$10$^{5}$), fast switching
time (2 $\mu$s, test limit), durable cycles demonstrated show
promising memory properties. The intermediate states observed
reveal the filamentary conduction nature.

\end{abstract}

\maketitle

\input epsf.sty \flushbottom

Resistive switching in various materials such as metal
oxides\cite{szot,yang}, chalcogenides\cite{hirose} and organic
materials\cite{ouyang,scot,weitz} has been intensively studied as
candidates for future nonvolatile memories\cite{waser}. Recently,
it also extends to new quasi-one dimensional (1D) and 2D materials
such as encapsulated nanowires\cite{cui}, multi-walled carbon
nanotubes (MWNTs)\cite{desh}, and graphene sheets\cite{st}. In
these nanostructures, the constriction in one dimension but less
in the other(s) facilitates the observation of the switching
events. The direct observations of nanosized gap or void
structures in these systems come to similar switching mechanisms
attributing to the electric close-and-break motion of the material
at the gap/void region. Less attention has been paid on the
substrate material of SiO$_{2}$ due to its good dielectric
(insulating) property. Meanwhile, the amorphous form of
SiO\cite{morgan,simmons,dear,dear2,thurs,shatzkes} or a defected
SiO$_{x}$ surface\cite{yao} can exhibit memory phenomena, in which
structural defects induced by high local field is one of the
proposed causes\cite{morgan}. For a gap system at nano size, it is
expected that a high local field is built up during the switches
between high-impedance (OFF) and low-impedance (ON) states. It
therefore carries the significance to investigate the local field
effect on the commonly used substrate material of SiO$_{2}$. In
the following context, we demonstrate resistive switching
phenomena in various nanogap systems on SiO$_{2}$ substrate, made
by different materials and means. The similar switching
characteristics in all the systems point to the most likely cause:
SiO$_{2}$ breakdown (BD) induced filaments, possibly through Si-Si
wire formation.

\begin{figure}[tbp]
\includegraphics*[width=8.6cm]{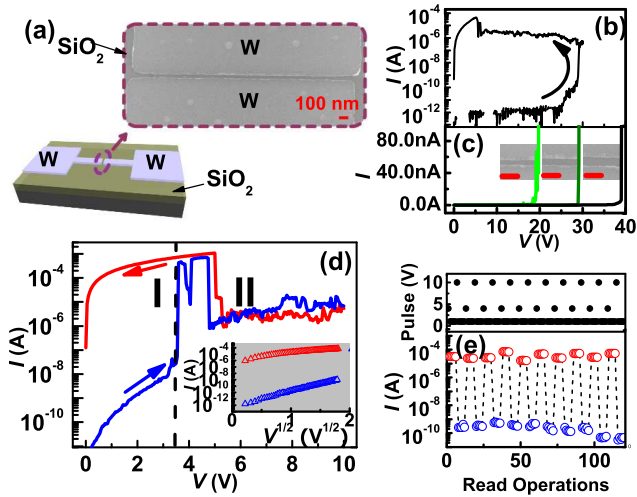}
\caption{(a). Schematic of the W-W gap and the SEM image. (b). IV
of the initial sweep from 0 V $\rightarrow$ 30 V $\rightarrow$ 0 V
in the as-made device. (c). IVs in three as-made devices with W-W
spacings of 30 nm (green curve), 50 nm (dark green curve), and 70
nm (black curve). The picture beside each curve shows the
corresponding SEM images, with the red scale bars indicate 100 nm.
(d). IVs of a forward (0 V $\rightarrow$ 10 V, red curve) and
subsequent backward (10 V $\rightarrow$ 0 V, blue curve) in the
electroformed device. The black dashed vertical line separates
region "I" (reading) and region "II" (writing/erasing). The Inset
shows the IVs in region "I" using a I-V$^{1/2}$ plot. (e). Memory
cycles of the device: after every five readings (+1V), the device
was set by an erasing pulse (+10 V) or a writing pulse (+4 V). The
top panel shows the corresponding pulses.}
\end{figure}

Shown in Fig. 1a, an initial gap system is a pair of tungsten (W)
electrodes separated by $\sim$50 nanometers (nm) on a
thermal-oxidized Si surface (the SiO$_{2}$ thickness is 200 nm,
and same thickness is used for all the following devices), defined
by standard electron beam lithography (EBL) and lift-off process.
Electrical characterizations were performed using an Agilent 4155C
semiconductor parameter analyzer in vacuum environment ($\sim$
10$^{-2}$ mTorr). Bias voltage was applied between the two
electrodes by sweeping from 0 V to 30 V and then back to 0 V (see
Fig. 1b). The device shows no conduction during the initial
forward sweep from 0 V to 25 V (e.g., the current is at the noise
level of the instrumentation of $\sim$ 10$^{-12}$ A). Substantial
conduction begins at $\sim$25 V with a sudden current rise at
$\sim$30 V, indicating a SiO$_{2}$ breakdown (BD). An irreversible
resistance change takes place in the post-BD device, indicated by
the increased current level during the subsequent backward sweep
from 30 V to $\sim$6 V. The sudden current (or conductance) rise
at $\sim$6 V in this backward sweep indicates the initiation of
hysteretic current-voltage curves (\textit{I-Vs}) essential for
memory switching. Fig. 1d shows the two characteristic
\textit{I-Vs} of the post-BD device: In a forward sweep (0 V
$\rightarrow$ 10 V, blue curve), beginning with an OFF
state\cite{star}, the device jumps to an ON state at $\sim$3.5 V
and goes back to OFF at $\sim$5 V; in the backward sweep (10 V
$\rightarrow$ 0 V, red curve), it jumps from an OFF state to an ON
state at $\sim$5 V and keeps the ON state below 5 V. Consequently,
a current hysteresis is produced in the bias range below 3.5 V
(region "\textbf{I}" in Fig. 1d). The underlying information about
the two \textit{I-Vs} is that a fast voltage drop edge above 3.5 V
(region "\textbf{II}" in Fig. 1d) can set the conductance of the
device into a value corresponding to that set
voltage\cite{morgan}. For example, a +4 V pulse "writes" the
device into an ON state, while a +10 V pulse "erases" the device
to an OFF state. The set states can be read out in the lower bias
region \textbf{I} without being destroyed, featuring the
nonvolatile memory property. Fig. 1e shows the corresponding
memory cycles by pulses as narrow as 2 $\mu$s (our instrumentation
limit) in the device, with an ON/OFF ratio approaching 10$^{5}$.

\begin{figure}[tbp]
\includegraphics*[width=8.6cm]{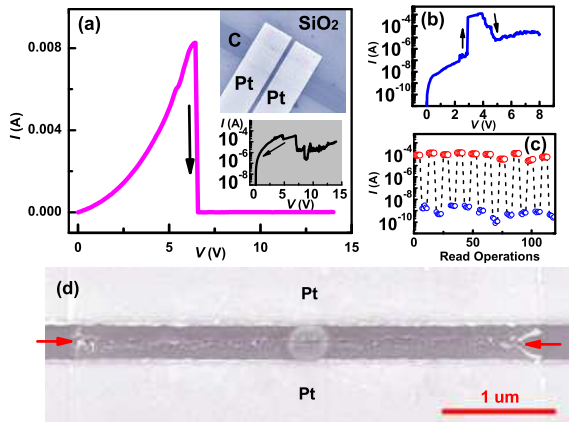}
\caption{(a). IV of the initial forward sweep in the as-made
$\alpha$-C device. Top Inset shows the SEM image of a patterned
device. The bottom Inset shows the IV of the subsequent backward
sweep right after $\alpha$-C BD. (b). The characteristic forward
IV in the electroformed device. (c). Memory cycles using +1 V (5
reads), +4 V (write), and +10 V (erase) pulses. (d). SEM image of
the $\alpha$-C stripe between the two Pt electrodes after the
$\alpha$-C BD. The red arrows indicate the BD induced gap region.}
\end{figure}

The SiO$_{2}$ BD induced conduction is supported by the linear
dependence of BD threshold voltage on the electrode-electrode
spacing. Gap spacings of $\sim$30, $\sim$50, and $\sim$70 nm
result in BD threshold values of $\sim$18 V, $\sim$29 V, and
$\sim$39 V, respectively (see Fig. 1c). The corresponding averaged
electric field is $\sim$6 MV/cm, which falls into the typical BD
values of SiO$_{2}$\cite{find} (surface region is also expected to
induce BD more easily than bulk given the less likeliness of
defect free). The sudden current increase during the first sweep
is accompanied by observable SiO$_{2}$ substrate damage in the gap
region. Subsequent forward or backward sweeps, usually undergo
gradual current increases and fluctuations having the
characteristics more and more alike those of the forward or
backward \textit{I-Vs} depicted in Fig. 1d. This electroforming
precess, resembles that observed in vertical M/SiO/M ("M" denotes
conducting electrodes) switching systems\cite{morgan}, in which
the amorphous form of SiO is the conducting and switching medium.
The non-ohmic \textit{I-Vs}, both for ON and OFF states, are
dominated by Poole-Frenkel conduction having the characteristic of
$\log(\emph{I}) \propto \emph{V}^{1/2}$ (see Inset in Fig. 1d).
The calculated\cite{cal} Poole-Frenkel field-lowering coefficient
($\beta_{PF}=3.4\times10^{-5} eV m^{1/2} V^{-1/2}$) from the OFF
state is very close to the theoretical one of $3.8\times10^{-5} eV
m^{1/2} V^{-1/2}$ and other experimental values in
SiO$_{x}$\cite{gould}.

\begin{figure}[tbp]
\includegraphics*[width=8.6cm]{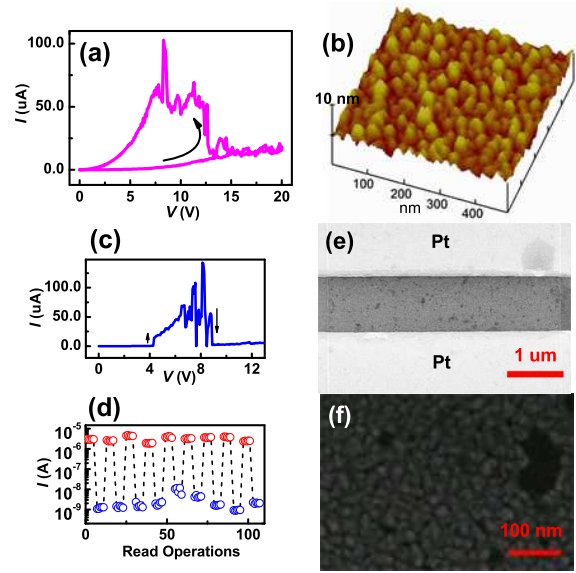}
\caption{(a). IV of the initial sweep (0 V $\rightarrow$ 20 V
$\rightarrow$ 0 V) in the as-made Al-island device. (b). AFM
surface morphology of Al film between the two Pt electrodes. (c).
The characteristic IV of a forward sweep in the electroformed
Al-island device. (d). Memory cycles using +1 V (5 reads), +6 V
(write), and +14 V (erase) pulses. (e). SEM image of the Al-island
after switching, showing no apparent gap structure. (f). SEM image
of the Al film of the switching device, showing grained
structures.}
\end{figure}

The electric field assisted BD in conducting
materials\cite{st,naitoh,tour} offers another means for gap
generation. A lift-off process was used to define an amorphous
carbon ($\alpha$-C) stripe ($\sim$50 nm thick, by sputtering from
a carbon graphite target) on SiO$_{2}$ substrate. Two platinum
(Pt) electrodes, with a comparatively large spacing ($\sim$0.4
$\mu$m), were then defined (see top Inset in Fig. 2a). Bias
voltage was applied between the two electrodes. The sudden current
drop at $\sim$6.5 V (see Fig. 2a) indicates a BD in the $\alpha$-C
stripe. Corresponding scanning electron microscope (SEM) image
reveals a cracked region perpendicular to the current direction in
the $\alpha$-C stripe (see Fig. 2d). The reduced conduction right
after the $\alpha$-C BD (see the subsequent backward sweep in
bottom Inset in Fig. 2a) has similar Poole-Frenkel feature,
indicating the disruption of the $\alpha$-C layer and simultaneous
BD in SiO$_{2}$ in the gap region. The conductance jump at $\sim$7
V (bottom Inset in Fig. 2a) during this backward sweep, initiates
the similar electroforming process as discussed above in the W-W
gap (see Fig. 1b). The characteristic forward \textit{I-V} (Fig.
2b) and switching (Fig. 2c) show similar features to those in the
W-W gap (Fig. 1d-e) such as current levels, writing/erasing
voltages, ON/OFF ratio, and switching time. While the threshold BD
voltage in $\alpha$-C tends to be proportional to the
electrode-electrode spacing, the writing/erasing voltages for
switching tend to be independent of it, consistent with the local
switching nature within the gap region (consider that the
collective resistance of contacts and $\alpha$-C layer is
considerably smaller than that of the gap region, the bias voltage
largely drops across the gap).

\begin{figure}[tbp]
\includegraphics*[width=8.6cm]{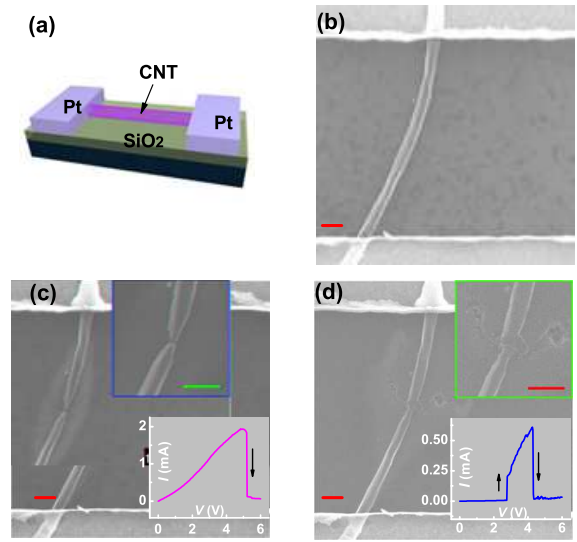}
\caption{(a). Schematic of a carbon nanotube patterned between two
Pt electrodes on SiO$_{2}$ substrate. (b) SEM image of the as-made
MWNT device before electrical characterization. (c). SEM image of
the same MWNT right after BD. The bottom Inset shows the BD IV and
the top Inset is a magnified view, showing gap structure at the BD
region. (d). SEM image of the same MWNT device after
electroforming. The bottom Inset shows the corresponding
characteristic forward IV and the top Inset is the magnified view
of the gap region, showing SiO$_{2}$ damage.}
\end{figure}

Compared to the EBL defined W-W gap, the $\alpha$-C BD induced gap
reduces the initial BD and electroforming voltage of SiO$_{2}$ in
the gap region, since the narrowest part is expected to be smaller
than, e.g., 30 nm. On the other hand, it also offers an better way
to investigate the details of the gap region by removing the
$\alpha$-C layer without destruction to the SiO$_{2}$ substrate.
Ultra-violet ozone exposure ($\sim$4 hours) was used to remove the
$\alpha$-C in the same switching device. SEM image shows
substantial damage to the SiO$_{2}$ part corresponding to the gap
region (compare Fig. 2d and e). Further control tests were
performed in devices with same $\alpha$-C thicknesses and
electrode spacings to investigate how the damage to SiO$_{2}$
forms. In one group, we produced BD in the $\alpha$-C layer by one
single sweep to a voltage above the $\alpha$-C BD threshold value,
while in the other we performed multiple sweeps up to a voltage
slightly below the $\alpha$-C BD threshold value (thus no gap
generation). After the $\alpha$-C was removed, observable damage
to the SiO$_{2}$ substrate at the gap region in the first group
was found, compared to no damage to the SiO$_{2}$ substrate in the
second group. The results reveal that: 1) gap generation in the
above $\alpha$-C layer simultaneously induces SiO$_{2}$ BD within
the gap region, which is consistent with the reduced conduction
(through post-BD SiO$_{2}$) having the Poole-Frenkel feature right
after the $\alpha$-C BD discussed above; 2) the damage to
SiO$_{2}$ is mainly through local electric-field induced BD, as
opposed to local heating, since a great reduction in current local
heating is expected after the disruption of $\alpha$-C layer in
the gap region due to the sudden current drop. We also found that
electroformed and switching devices tend to have more apparent
damage in the SiO$_{2}$ substrate than the ones having initial
$\alpha$-C BD. These results indicate the role of SiO$_{2}$ in
switching in the gap region.

\begin{figure}[tbp]
\includegraphics*[width=8.6cm]{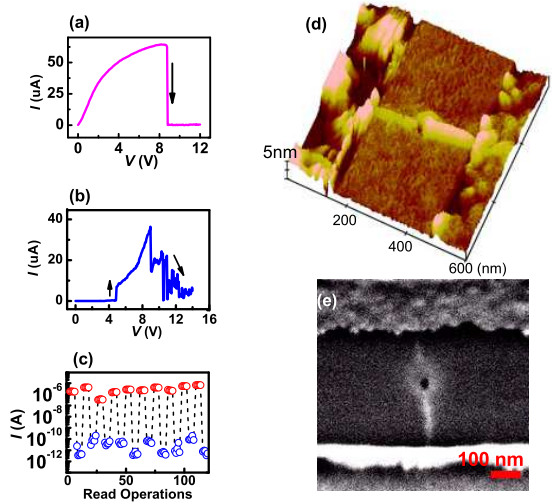}
\caption{(a). The initial BD IV of the SWNT device. (b). The
characteristic forward IV in the electroformed device. (c). Memory
cycles using +2 V (5 reads), +8 V (write), and +14 V (erase)
pulses. (d). AFM image of the electroformed SWNT device. (e).
Corresponding SEM image of the same electroformed SWNT device.}
\end{figure}

The post-BD SiO$_{2}$ switching nature is further emphasized by
using a different material as the gap generation medium.
Electrical BD in a titanium nitride (TiN) stripe on SiO$_{2}$
substrate leads to similar gap structure and switching (see
supplementary material Fig. 7 at the end of the article). Compared
to that in $\alpha$-C stripe, the gap in a TiN stripe is usually
located at the TiN-electrode interface, instead of in between the
electrodes. The possible reason is that the Schottky barrier at
the TiN-electrode interface (since TiN is semiconducting) enhances
the local field and facilitates the TiN BD there, as opposed to a
good C-Pt electric contact in the $\alpha$-C stripe, where C BD is
likely to happen at the least heat-dissipation region far away
from both electrodes.

\begin{figure}[tbp]
\includegraphics*[width=8.6cm]{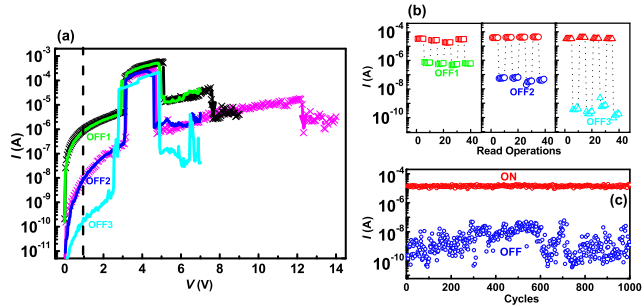}
\caption{(a). IV evolutions in an electroformed MWNT device:
starting from an initially established forward IV (0 V
$\rightarrow$ 7 V, green curve), the subsequent forward sweep
(black crossed curve) up to a higher voltage (0 V $\rightarrow$ 9
V) lowers OFF state (see black arrow at $\sim$7.5 V). A new
characteristic forward IV with a lower OFF state establishes
subsequently (blue curve). By sweeping to an even higher voltage
(magenta crossed curve), a second conductance reduction in OFF
state is initiated (see magenta arrow at $\sim$12 V). Similarly, a
third characteristic IV featuring even lower OFF state establishes
thereafter (cyan curve). (b). Memory cycles using the same set of
reading/writing/erasing pulses of +1/+3.5/+7 V in the same device,
with the left, middle and right columns corresponding to the
established green, blue, and cyan characteristic IV curves in (a).
(c). A 10$^{3}$ cycles in a second MWNT device.}
\end{figure}

Gap by electric-field BD is less likely to form in metal stripes
because the high current density usually melts the metal before a
BD and the surface tension of liquid metal tends to form droplets,
preventing a narrow and well aligned gap. However, thin metal film
tends to form discreet islands\cite{metalislands}, and nano gaps
may form naturally between individual islands. For this purpose,
aluminum (Al) thin film ($\sim$10 nm thick) was deposited by
sputtering on SiO$_{2}$ between two Pt electrodes. The surface
morphology of the deposited Al studied by atomic force microscope
(AFM) shows discontinuous grained feature (Fig. 3b). Voltage sweep
was applied between the two electrodes for the as-made device.
Unlike that in $\alpha$-C or TiN stripes, the initial forward
sweep (0 V $\rightarrow$ 20 V) here shows much lower conductance
and no sudden current drop (see Fig. 3a). This further indicates
the discontinuity of the Al film. The subsequent backward sweep
(20 V$\rightarrow$ 0 V), with similar conductance jump, indicates
the initiation of hysteretic behavior. The SEM image of the
electroformed device shows grained Al surface between the
electrodes (Fig. 3f) with no such apparent gap (Fig. 3e) as that
produced by BD in $\alpha$-C or TiN stripes, which is consistent
with the \textit{I-V} feature of the initial sweep discussed and
in support of the idea of switching in as-formed island-island
gaps. Although the actual switching site is unknown due to
numerous indistinguishable gaps between islands, it is expected
that the relatively high resistance of Al film reduces both the
current and effective voltage drop across the switching site.
Therefore, the switching device has lower ON current (Fig. 3d) and
higher writing/erasing voltages (Fig. 3c).

From the device part, one interesting question is how small the
device can go. On the other hand, a small and constricted
switching size may offer clearer view of the switching event,
while it is relatively difficult to distinguish the actual
switching site in a wide stripe, e.g., whether the switching
happens uniformly along the entire gap region or locally. The
electrical BD in MWNTs\cite{chiu,huang} provides a potential means
for further reduction in gap size. For this purpose, a MWNT with
diameter $\sim$60 nm was patterned between two Pt electrodes (see
Fig. 4a for illustration). Fig. 4b shows the SEM image of the
pristine device. Electrical BD begins at a voltage $\sim$5 V
indicated by a sudden current drop (see bottom Inset in Fig. 4c).
The corresponding SEM image right after this BD shows a broken gap
region (see Fig. 4c and top Inset). The subsequent sweeps
electroform the device, showing the characteristic switching
\textit{I-V} (bottom Inset in Fig. 4d) similar to those observed
in all the above devices. The SEM image of the electroformed
device shows clear damage to the SiO$_{2}$ at the gap region,
extending outside (see Fig. 4d and top Inset). It is necessary to
point out that switching in electrical BD MWNTs was reported
before\cite{desh}. The ON and OFF states were achieved by
close-and-break motion of the carbon nanotube shells from the two
broken ends, and were stable up to only several cycles\cite{desh}.
The switching here is attributed to the post-BD SiO$_{2}$ at the
gap region and can be stable against cycling (see Fig. 5c). The
non-mechanical switching is also supported by the high yield in
our devices (e.g. 10 out 10 MWNTs tested show similar switching),
regardless of the actual details of the broken ends. This is in
contradictory to mechanical switching as we expect both the broken
gap size and morphology of broken ends would affect the switching.

Single-walled carbon nanotubes (SWNTs) provide the candidates for
ultra-small constrictions. Electrical BD in metallic SWNTs was
reported and used as a means for semiconducting SWNTs
sorting\cite{avouris}. Metallic SWNT with a diameter $\sim$2 nm
was patterned between two Ti/Pt electrodes. Electrical BD takes
place at $\sim$8.5 V (see Fig. 5a). Fig. 5b-c show the
characteristic \textit{I-V} and switching after BD and
electroforming. The AFM image of the electroformed device shows
broken region in SWNT (Fig. 5d). The corresponding SEM image in
Fig. 5e shows a dark dot at the gap region, indicating hole-like
damage to the SiO$_{2}$ substrate, which is also inferred from the
AFM image. The comparatively small ON current (see Fig. 5c) in the
SWNT device is mainly attributed to the contact resistance between
the broken nanotube ends and the post-BD SiO$_{2}$ in the gap
region, as a good electric contact to metallic SWNTs with diameter
below 2 nm usually requires specific metals\cite{dai2}. This
contact resistance also reduces the effective voltage drop across
the gap, resulting in higher writing/erasing voltages (see Fig.
5b).

All the above devices share the similarities once operational.
These include similar writing/erasing voltages and currents (the
reduced ON currents and increased writing/erasing voltages in
Al-island and SWNT devices are due to high film or contact
resistances), switching time, and noise distributions. The
comparatively large current fluctuation in the erasing bias region
observed in all the devices is another characteristic of SiO$_{x}$
conduction\cite{morgan}. The largely independence of switching on
the (effective) electrode materials ranging from metal (W, Al),
metallic nonmetal (C), semiconductor (TiN), to carbon nanotubes,
along with the electroforming processes and observable damage to
the SiO$_{2}$ substrate in the gap regions, bespeaks the intrinsic
post-BD SiO$_{2}$ switching nature. This is confirmed by making
similar stripes and nanotube structures on Si$_{3}$N$_{4}$
substrate, in which switching was not observed after the gap
generation (severe substrate damage is usually observed in the gap
region by an attempt of electroforming). Another common feature
observed in all the above devices is the various intermediate
conduction states. As shown in Fig. 6a, for an initially
established characteristic forward (0 $\rightarrow$ 7 V)
\textit{I-V} (green curve), an OFF state with a current level of
$\sim$ 10$^{-6}$ A can be set by a +7 V pulse (see cycles in the
left column in Fig. 6b). By sweeping to a higher voltage above 7 V
(black crossed curve), a lower conduction state appears,
signatured by a sudden current drop at $\sim$7.5 V (see black
arrow). In the subsequent sweep, a new \textit{I-V} featuring a
lower OFF state establishes (see blue curve), and the same +7 V
pulse now sets the OFF state to a current level of $\sim$
10$^{-8}$ A (see cycles in the middle column in Fig. 6b). The OFF
current can be further reduced by sweeping to an even higher
voltage (see magenta crossed curve) during which a second
conduction reduction appears at $\sim$12 V (magenta arrow). This
leads to a third \textit{I-V} (cyan curve) with an OFF current
level of $\sim$ 10$^{-10}$ A that can be set by the same erasing
pulse of +7 V (see cycles in the right column in Fig. 6b).
Multiple intermediate conduction states is an indication of
filamentary conduction in oxides\cite{kim}, in which different
states can be viewed as formation/termination of new/existent
percolation paths\cite{kim,chen}. To an extent, the current
fluctuations in the erasing region (see cyan curve in Fig. 6a) can
be viewed as various meta-stable states or excitons, e.g., same
erasing voltages can produce different OFF currents if they
encounter current fluctuations of different magnitudes. This is
well reflected in a 10$^{3}$ cycles, in which the OFF states
undergo various conductances (see Fig. 6c). We attribute this to
be the main cause of instability in the current device
performance.

In summary, we have demonstrated reproducible memory switching in
various nanogap systems on SiO$_{2}$ substrates. The lack of
dependence of the switching behaviors on electrode materials
points to a common mechanism, post-BD SiO$_{2}$ switching in the
gap region. It is therefore important to exercise caution when
building resistive switching nanosystems on SiO$_{2}$ substrate.
Effects should be taken to distinguish the switching cause. The
high ON/OFF ratio, fast switching time, and durable cycles
demonstrated here show interesting memory properties. In
particular, the small switching site demonstrated in a SWCNT shows
the feasibility of high-density SiO$_{2}$-based memory arrays if a
vertical embodiment could be realized. The observed intermediate
states reveal the filamentary conduction nature in post-BD
SiO$_{2}$ switching which is likely Si-Si wire formation, although
a further investigation of the individual filamentary path is
needed. The post-BD SiO$_{2}$ conduction suggests another possible
mechanistic scenario for the switching that was observed in our
graphitic nanocables\cite{tour}.

\begin{figure}[tbp]
\includegraphics*[width=6cm]{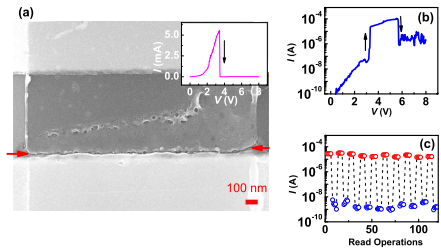}
\caption{Supporting material (a). SEM image of a switching TiN
stripe on SiO$_{2}$ substrate. The red arrows indicate the gap
region induced by an initial TiN BD. The Inset shows the initial
TiN BD I-V. (b). The characteristic forward I-V in the
electroformed TiN device. (c). Memory cycles using +1 V (5 reads),
+4 V (write), and +7 V (erase) pulses.}
\end{figure}


\begin{thebibliography}{*}
\bibitem{star} {A fast voltage drop edge\cite{morgan} is expected at
the end of each forward sweep, which sets the device state. So the
beginning conduction state for a forward sweep is determined by
the previous forward sweep. Here the characteristic forward IV was
obtained after at least another forward sweep prior to it in the
same voltage range. So it begins with an OFF state. All the
following characteristic forward IVs were obtained in this way,
unless otherwise specified.}
\bibitem{cal} {The Poole-Frenkel expression\cite{gould} is
$I=I_{0}exp(\frac{\beta_{PF}E^{1/2}}{kT})$, where $k$ is the
Boltzmann constant, $T$ the temperature, and $E$ the electric
field. We used the room temperature and a gap width $d=50$ nm
($E=V/d$) for fitting.}
\end{thebibliography}
\end{document}